\documentstyle[psfig,aps,prb,twocolumn,floats]{revtex}

\begin{document}
\twocolumn[\hsize\textwidth%
\columnwidth\hsize\csname@twocolumnfalse\endcsname

\title {Collective Modes and Electronic Spectral Function in Smooth
Edges\\
of Quantum Hall Systems} 
\author{Sergio Conti}
\address{Scuola Normale Superiore, 56126 Pisa, Italy}
\address{Department of Physics,  University of
Missouri, Columbia,  Missouri 65211}
\author{Giovanni Vignale} 
\address {Department of Physics,  University
of Missouri, Columbia,  Missouri 65211}
 
\date{August 28, 1996} 
\maketitle

\begin{abstract}
We present a microscopic theory of the collective modes of a
``smooth'' 
edge of a quantum Hall system, showing under what conditions these
modes can be described as a set of independent bosons.  We then
calculate 
the electronic spectral function in an independent-boson model - a
procedure that reduces to standard bosonization in the limit of
``sharp'' edge.  The I-V tunneling characteristics deduced from this
model exhibit, for low voltage, a power law behavior, with exponents
that differ significantly from those of the sharp edge model. 
\end{abstract}

\vskip1pc]
\narrowtext

Understanding the character of edge excitations is crucial to the
theory of the transport properties of two dimensional systems in the
presence of a strong perpendicular magnetic field, such as quantum
Hall 
bars, quantum wires and dots \cite {general}.
An effective theory of edge excitations was first derived by 
Wen \cite{Wen}.  He showed that  a ``sharp'' edge (see below) is
 a realization of the one-dimensional chiral Luttinger liquid (CLL)
model, where the electronic spectral function exhibits, in general, a
non 
trivial behavior, leading to a density of states that vanishes, at
low energy, as a power law.  This theory has been well confirmed by
detailed microscopic calculations \cite{Palacios} and by recent
experiments \cite{Webb,Chang}.  The effect of the long range of the
Coulomb interaction, which was initially ignored, has been recently
included by Z\"ulicke and MacDonald \cite{Zulicke}. 

All the above papers assumed the validity of the so-called ``sharp
edge''  model, in which the density of the system drops sharply from
the bulk value $\rho_0$ to zero within a few magnetic lengths
$l = (\hbar c /eB)^{1/2}$.  There are numerous indications that this
is 
not always the correct model for the edge.  On one hand, Hartree-Fock
calculations \cite {Chamon} for strongly confined systems,  predict
that, at sufficiently strong magnetic field, the edge  undergoes a
reconstruction, taking up a more extended shape.  On the other hand,
in the case of {\it 
smooth} confinement, such as can be realized by gate electrodes, the
electronic density is expected to have a smoothly varying profile (on
the scale of $l$), determined by classical electrostatic equilibrium
\cite{Beenaker,Chklovskii}. Detailed calculations using density
functional theory, 
Thomas-Fermi theory, and other methods
\cite{Ferconi,Heinonen,Brey,Dmitri,Gerritt} have  confirmed the
theoretical validity of this ``smooth edge" picture.  Edge imaging
experiments \cite {Haren}  have confirmed the relevancy of this 
description for gate-confined Hall bars.  

In this Communication we want to investigate the spectral properties
of  the ``smooth edge'' model described above.  The mapping to a one
dimensional chiral electron liquid is not justified in this case. In
fact, a recent study by Aleiner and Glazman (AG)\cite{Aleiner}, based
on the classical hydrodynamics approach, has shown that a 
smooth edge, in contrast to a sharp edge,  supports {\it multiple
branches} of edge waves.  Of these, one is the usual edge
magnetoplasmon \cite{Volkov}, and 
the others (infinitely many in the classical approach of AG) are
phonon-like and lower in energy than the magnetoplasmon. 
We shall show
that, under the assumption of smooth density variation, only a finite
number of these phonon-like modes are correctly described as
independent bosons.

In order to calculate the electronic spectral function we rely on the
strong analogy between this problem and that of a uniform 
electron gas in the 
partially filled lowest Landau level (LLL).  The  similarity  between
these two 
problems arises from the fact that in both cases, in  a mean field
approximation, the self-consistent potential is uniform, so that the
electrons 
are distributed among a large number of degenerate orbitals at the
Fermi energy. 
In  a smooth edge this  occurs because the nonuniform
electronic density  perfectly screens the field  due to the external
confinement potential\cite{Beenaker,Chklovskii}. Recently, Johansson
and 
Kinaret \cite{Kinaret} have shown that a qualitatively correct
description of 
the spectral function \cite{He} of the uniform electron 
gas in the LLL at 
general filling factor is given by an independent boson model
(IBM)\cite 
{Mahan} in which a single localized electron interacts with the
density 
fluctuations of the system. An essentially equivalent procedure has
been applied by Aleiner, Baranger and Glazman\cite{ABG} to study the
spectral 
function of the two-dimensional electron liquid in a weak magnetic
field. Finally, a formal justification of the IBM from diagrammatic
many-body 
theory  has been provided by Haussmann\cite{Haussmann}.  
Encouraged by these 
successes, here we  apply the independent boson model to the  problem
of the 
smooth edge.  The resulting theory reduces to standard bosonization 
in the limit 
of a sharp edge, i.e.,  when there is only one branch of edge waves.
When multiple branches are present, our results for the low energy
behavior of the 
tunneling density of states are significantly different from those of
the sharp 
edge model.  The actual number of modes that must be included depends
on the width of the edge, as explained below.

Let us begin by writing down the microscopic Hamiltonian, within the
lowest Landau level, in terms of density fluctuations relative to
the equilibrium density profile $\rho_0(y)$:
\begin {equation}
H = {1 \over 2} \int_{edge} {e^2 \over \vert \vec r - \vec r' \vert}
\delta \rho 
(\vec r)  \delta \rho (\vec r') d^2r d^2r',
\label {H}
\end {equation}
where the density operator (projected  in the LLL) has been 
written as
\begin {equation}
\rho (\vec r) = \rho_0(y) +\delta \rho (\vec r).
\label{ro}
\end {equation}
The integral in eq.~(\ref{H}) extends over the edge region which we
take to be  
$0<x<L$, $0<y<d$, with $L\gg d\gg l$, and
translationally invariant along $x$ ($\rho_0(y)=0$ for $y<0$).
The projected density fluctuation operator is given by
\begin {equation}
\delta \rho (\vec r) = {1 \over \sqrt{\pi} l L} \sum_{h \neq k}
c_k^\dagger c^{}_h e^{i(h-k)x} e^{-{(y-y_k)^2+(y-y_h)^2 
\over 2 l^2}},
\label{deltaro}   
\end {equation}
 where $y_h=hl^2$, $h$ and $k$ are  integral multiples of $2 \pi/L$,
$c_k^\dagger$ is the creation operator of a Landau gauge orbital
centered a $y_k$ with wave vector $k$ in the $x$ direction.  Note the
restriction $h \neq k$ which excludes the equilibrium component 
of the density.
The kinetic energy is absent in eq.~(\ref{H})
due to projection on the LLL, and we have assumed
that density fluctuations vanish in the bulk of the system, 
i.e., the bulk
is incompressible.  The terms linear in $\delta \rho$  have 
vanished because of
the equilibrium condition $\int \rho_0(\vec r) v(\vec r - 
\vec r')d^2r'
+V_{ext}(\vec r) = constant$ where $V_{ext}(\vec r)$ is the 
confinement
potential.  Therefore the problem is formally  similar to that of a
translationally invariant electron gas:  the nonuniformity 
enters only through 
the  restricted region of integration in eq.~(\ref{H}). 

The normal mode operators $\delta \rho_{nk}$ are now
introduced according to the definition
\begin {equation}
\delta \rho_{nk} = \int_0^L {dx \over L} e^{-ikx} \int_{0}^d dy 
f_{nk}(y)\delta \rho (x,y),
\label {deltaronk}
\end{equation}
where $f_{nk}(y)$ are the  solutions of the equation
\begin {equation}
\int_{0}^d K_0(k \vert y-y' \vert)f_{nk}(y'){\rho_0'(y)
\over\bar\rho}dy' = 
{1 \over \lambda_{nk}} f_{nk}(y),
\label {eigenvalueproblem}
\end {equation}
where $K_0(y)$ is the modified Bessel function.  
They satisfy the
orthonormality condition 
$\int_{0}^d f_{nk}(y)f_{mk}(y){\rho_0'(y)
\over \bar \rho}dy = \delta_{nm}$, and vanish outside the interval 
$[0,d]$.
Equation ~(\ref{eigenvalueproblem}) is  the eigenvalue
problem solved by AG.  The $n$-th eigenfunction has $n$ nodes 
in the $y$
direction.  In terms of the normal modes, the hamiltonian (\ref{H})
takes the form
\begin{equation}
H = \sum_{nk>0} \hbar \omega_{nk} b_{nk}^\dagger b^{}_{nk},
\label {Hnormal}
\end {equation}
where the  operators $b_{nk}$ are defined via $\delta \rho_{nk}
\equiv \sqrt{k l^2 \bar \rho/L} b_{nk}^\dagger$, and the
eigenfrequencies $\omega_{nk}$ are  given by $k \bar \nu 
e^2/\lambda_{nk}\pi$,
and $\bar \nu = 2 \pi l^2 \bar \rho$ is the usual filling 
factor in the bulk.

It remains to be determined under what conditions the operators
$b_{nk}$ are good boson operators.  To this end we 
substitute eq.~(\ref{deltaro}) in eq.~(\ref{deltaronk}), noting that 
when $n\ll d/l$ the gaussian factors in the integral 
can be replaced by $\delta$-functions on the scale of variation of
$f_{nk}$. We obtain
\begin {equation}
\delta \rho_{nk} \simeq e^{-k^2l^2/4} {1 \over L} \sum_h 
c_{h-k/2}^\dagger
c^{}_{h+k/2} f_{nk}(hl^2).
\label {deltaronkslow}
\end {equation}
$(n\ll d/l)$. 

The commutator of two density fluctuations can now be easily 
calculated to be
\begin {eqnarray}
[\delta \rho_{nk},\delta \rho_{m-k}] &=&    
 e^{-k^2l^2/2} {1 \over L^2}\sum_h (n_{h-k/2}-n_{h+k/2}) \times
 \nonumber\\
&&\times f_{nk}(hl^2) f_{mk}(hl^2)
\simeq -{kl^2 \bar \rho \over L} \delta_{nm},
\label {commutator}
\end {eqnarray}
in agreement with the commutation rules for bosons.  In arriving at
equation ~(\ref{commutator}) we have assumed $kl\ll 1$, and we have
replaced the occupation number operators by their ground-state
expectation values $n_k$, which amounts to a linearization of the
equations of motion around the equilibrium state.  For the 
commutator $[\delta
\rho_{nk},\delta \rho_{m-k'}]$, with $ k \neq k'$, we find, 
at the same level of
approximation, zero. This is because the commutator in 
question contains terms
of the form $c_h^\dagger c^{}_{h'}$ with $h \neq h'$, which 
vanish upon averaging
in a translationally invariant (along $x$) state.

Having thus completed the bosonization of the hamiltonian, we proceed
to the calculation of the spectral function within the independent
boson model\cite{Kinaret,Mahan}. The model describes a single
electron, localized at point $\vec r$, electrostatically coupled to
density 
fluctuations: 
\begin {eqnarray}
H_{IBM} &&= \sum_{nk>0} \hbar \omega_{nk} b_{nk}^\dagger b^{}_{nk}
+ \nonumber\\
&& +\psi^\dagger(\vec r) \psi(\vec r)  \sum_{nk>0} M_{nk}(y)
 [b_{nk}^\dagger e^{ikx}+b^{}_{nk}e^{-ikx}],
\label{Hindependentbosons}
\end{eqnarray}
where the matrix element $M_{nk}(y)$ is given by
\begin {equation}
M_{nk}(y) = {2e^2 \over \lambda_{nk}}f_{nk}(y)\sqrt{k\l^2
\bar\rho\over L},
\label{matrixelement}
\end {equation}
and $\psi^\dagger(\vec r)$ is the field operator that creates an electron in
the LLL coherent state (gaussian) orbital centered at $\vec r$.  The hamiltonian
(\ref{Hindependentbosons}) can be solved by standard methods \cite{Mahan},
within the one-electron Hilbert space.  The fermionic Green's function is
obtained as \begin{eqnarray}
&&G_>(y;t) \equiv -i \langle \psi(\vec r,t)\psi^\dagger(\vec r,0)
\rangle \nonumber\\ 
&&= (1 -\nu_0(y))\exp \left ( \sum_{nk>0} {M_{nk}^2(y) \over
\omega_{nk}^2}[e^{-i\omega_{nk}t}-1] \right),
\label{Greensfunction}
\end{eqnarray}
where $\nu_0(y) \equiv 2 \pi l^2 \rho_0(y)$,  and the sum over 
$n$ and $k$ in the
exponent is restricted by the conditions $n\ll d/l$ and $k\ll 
1/l$, which define
the regime of validity of the hydrodynamic approximation. 
This result can also
be obtained from direct bosonization of the electron field 
operator, as in
\cite{ABG}. 
The Fourier
transform of $G(y,t)/2\pi$ is the spectral function 
$A_>(y,\omega)$ and
gives the local density of states, which controls the 
tunneling current from a
point contact located at position $y$ into the edge.
 From eq.~(\ref{Greensfunction}) it can be easily shown 
\cite{Minnhagen} that
$A_>(y,\omega)$ satisfies the integral equation
\begin {equation}
\omega A_>(y,\omega) = \int_0^\omega g(\Omega) A_>(y,\omega - \Omega)
d\Omega,   
\label {integralequation}
\end {equation}
where
\begin {equation}
g(y,\Omega) \equiv \sum_{nk} {M_{nk}(y)^2 \over \omega_{nk}} 
\delta(\Omega -
\omega_{nk}).
\label {gofomega}
\end {equation}
Eq.~(\ref{integralequation}), together with the conditions 
$A_>(y,\omega) = 0$ 
for $\omega <0$ and $\int_0^\infty A_>(y,\omega) d \omega =
1 - \nu_0(y)$, completely
determines the spectral function. This equation further implies 
that, at
sufficiently small $\omega$, $A_>(y,\omega)$ will have  a 
power-law behavior
\begin {equation}
A(y,\omega) \sim \omega^{g(y,0) -1}
\label {powerlaw}
\end {equation}
if and only if the function $g(y,\Omega)$ has a finite limit 
for $\Omega \to
0$. The tunneling current $I$, as a function of voltage $V$, 
will then behave
as $V^{g(y,0)}$ for sufficiently low voltage.  Notice that 
this conclusion is
completely general, and does not depend on the specific 
(hydrodynamic) model 
that led to the definition of $g(y,\Omega)$ in 
eq.~(\ref{gofomega}). In the
general case, $g(y,\Omega)$ could be computed from the microscopic
density-density response function
$\chi(\vec r,\vec r',\Omega)$ of the edge as
follows: 
\begin {equation}
g(y,\Omega) = \int d^2r' d^2r'' v(\vec r - \vec r')v(\vec r - \vec
r'') Im \chi(\vec r',\vec r'',\Omega)/\Omega,
\label {gmicro}
\end {equation}
where $v(\vec r - \vec r')$ has the Fourier transform $v(\vec k) 
= (2 \pi e^2 /k)
\exp {(- (kl)^2/4)}$ \cite{Haussmann}. An important advantage 
of this microscopic
formulation is that the finite lifetime of the collective modes 
(which is
assumed to be infinite in the hydrodynamic model)  would be 
taken into account
through the width of the peaks in $Im \chi$.

The calculation of the exponent $g(y,0)$ is easily performed 
within the
hydrodynamic model. Neglecting the weak nonlinearity of the 
$n=0$ mode
\cite{Footnote2} we obtain $g(y,0) = \sum_n \beta_n(y)$, where
\begin {equation} 
\beta_n(y) = {1 \over \bar \nu}f_{n0}^2(y). 
\label {exponent} 
\end {equation}

Although  the cutoff  at $n=d/l$ introduces an uncertainty in 
the evaluation of
the exponent at any given $d$, we emphasize that there would 
be no uncertainty if
one used the microscopic formula ~(\ref{gmicro}) for 
$g(y,\Omega)$.  Our
approximate hydrodynamic evaluation of the exponent
should be in good qualitative agreement with the results of  
the more accurate
microscopic calculation.

We observe that independently of the shape of the density profile
$\beta_0(y) = 1/ \bar \nu$, with negligible corrections
arising from the weak nonlinearity of the dispersion of the 
$n=0$ mode.
Therefore in the sharp edge limit, when only one branch of edge 
waves exists,
we recover the familiar result of Wen's theory $A_>(\omega) \sim
\omega^{1/\bar \nu-1}$.
For $n>1$ $\beta_n(y)$ fluctuates around an
average value of $1/\bar \nu$ in a way dependent
on the form of the equilibrium density profile of the edge.
This can be confirmed by explicit calculation in the special case 
$\rho_0(y) = (2/\pi) \bar \rho \mbox{ atn}\sqrt{y/d}$
considered by AG for a gate-confined electron gas, leading to 
the result
$\beta_n(y) = {1 \over \bar \nu}
 T_{2n}^2\left(\sqrt{d /( y+d) }\right)(2-\delta_{n0})$, 
where $T_n(y)$ is the $n$-th Chebyschev polynomial. 
We conclude 
that the exponent in eq.~(\ref{powerlaw})  increases 
linearly with $d$
and therefore that in
the limit $d \to \infty$ (limit of infinitely smooth edge) 
the tunneling
density of states vanishes at low energy faster than any power 
law,  that is,
a ``hard'' gap develops.  However,  it is easy to see that the 
power law
behavior of eq.~(\ref{powerlaw})  only holds for $\omega\ll 
\bar \nu e^2/d \pi$ -
an interval that shrinks to zero for $d \to \infty$.

\begin{figure}
\psfig{figure=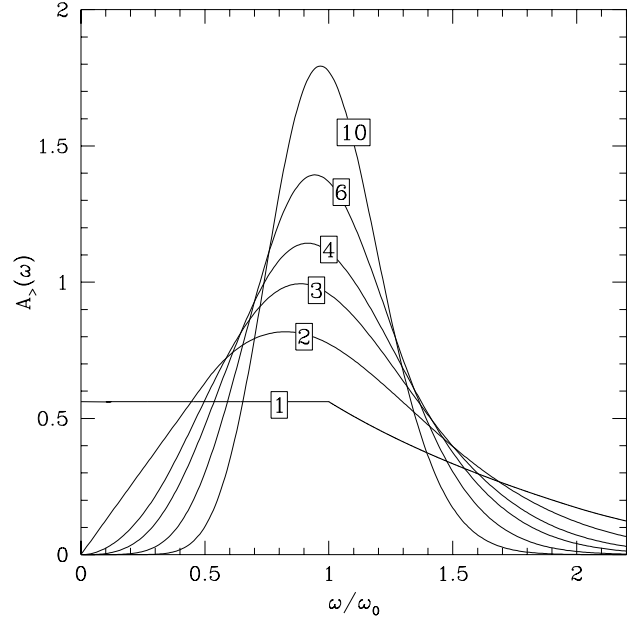,width=\columnwidth}
\caption{Electronic spectral function $A_>(\omega)$ as a function of
$\omega/\omega_0$, where $\omega_0=\bar \nu e^2/\pi l$, 
for edges of a
$\bar \nu=1$ QH system with 1, 2, 3, 4, 6 and 10 modes. Logarithmic
corrections to the edge magnetoplasmon dispersion are neglected.
The dependence on $y$ and $\rho_0(y)$ have been eliminated neglecting
the constant $1-\nu_0(y)$ and using $\beta_n=1/\bar \nu$,
$\omega_{nk}=\bar\nu e^2 k /\pi n$. The integral over $k$ has been 
cut off as explained in the text.} 
\label{figura}
\end{figure}

In Figure 1 we present our numerical results for the full electronic
spectral function, calculated from 
eq.~(\ref{integralequation}) within the
hydrodynamic model for different edge widths $d$.
In contrast to the analysis of the low-frequency
behavior, this calculation depends on the detailed form 
of the eigenfunctions
$f_{nk}(y)$ and eigenfrequencies $\omega_{nk}$.  
From a detailed  study of
the solutions of the eigenvalue equation 
(\ref{eigenvalueproblem}) we have found
that the $f_{nk}$'s  can be treated as being 
independent of $k$ and the
$\omega_{nk}$'s to be linear functions of $k$ 
up to a maximum wave vector $k_c =
n/d$  for which the wavelength along the edge equals the wavelength
perpendicular to the edge. For $k>k_c$ the mode 
dispersion becomes approximately
constant, and the wavefunction $f_{nk}$ 
becomes localized near the boundaries of
the edge region, giving negligible 
contribution to the spectral function.  The
results presented in Fig.~1 have been 
obtained using the double cutoff $n<d/l$
and $k<k_c$:  the results are found to be 
largely independent of the details of
the cutoff procedure.
	
Figure 1 shows clearly how the low energy pseudogap becomes
more and more pronounced with increasing $d$ (and, therefore,
increasing number of branches of 
edge waves).  For very large $d$ the spectral function 
is found to converge to
a $\delta$ function centered at $\omega_0 = \bar \nu 
e^2/\pi l$, which
coincides with the simplest estimate of the potential energy cost for
the insertion of an electron into a frozen liquid\cite{Haussmann}.

Our  results for $d\to\infty$  are in qualitative 
agreement with those
obtained in refs.~\onlinecite{Kinaret} and \onlinecite{Haussmann}  
for the spectral function of the
homogeneous electron gas, except that the latter is 
found to have a finite
width. This happens because our hydrodynamic approach 
is unable to give the
gapful collective modes of the homogeneous fluid phase\cite{Girvin}, 
and hence our spectral
function does not reduce to that of the homogeneous phase.

In conclusion, we have performed an independent boson
 model calculation
of the tunneling  density of states for a smooth edge, 
and we have found
that  it vanishes at low frequency as a power, with an exponent that
differs significantly from the one found in the sharp edge case.  
Recent experiments by Chang \cite {Chang} have 
apparently confirmed the
predictions of the CLL for the exponent of the tunneling density of
states in a sharp edge.  It should be interesting to extend these
studies to see if and how the exponents change as the smoothness of
the edge is varied.

 We gratefully acknowledge support from NSF
grant No. DMR-9403908. One of us (SC)  acknowledges support
from a  travelling scholarship from Scuola Normale Superiore. 
We thank Allan MacDonald and Ulrich Z\"ulicke for discussions
and for sharing the results of their work prior to publication.
One of us (GV)  acknowledges very useful discussions 
with I.~L.~Aleiner
and L.~I.~Glazman and the hospitality of the Aspen Center of Physics
where part of the work  has been done.


\begin{references}
\bibitem {Footnote2}  The $n=0$ mode has a dispersion of the form
$-k\ln k$.  Its deviation from linearity does not  
affect significantly the tunneling density of states.
\end{references}
\end{document}